\documentclass[twocolumn,preprintnumbers,prb]{revtex4}
\usepackage{amsfonts}
\usepackage[T1]{fontenc}
\usepackage{amsmath,amsbsy,amssymb,graphicx}
\usepackage{times}

\let\mathbf=\boldsymbol
\def\beginABC{\begin{subequations}}
\def\endABC{\end{subequations}}

\begin{document}

\title{ {\Large Spin-Valleytronics in Silicene:}\\
{\Large Quantum-Spin-Quantum-Anomalous Hall Insulators}\\
{\Large and Single-Valley Semimetals}}
\author{Motohiko Ezawa}
\affiliation{Department of Applied Physics, University of Tokyo, Hongo 7-3-1, 113-8656,
Japan }

\begin{abstract}
Valley-based electronics, known as valleytronics, is one of the keys to
break through to a new stage of electronics. The valley degree of freedom is
ubiquitous in the honeycomb lattice system. The honeycomb lattice structure
of silicon called silicene is an fascinating playground of valleytronics. We
investigate topological phases of silicene by introducing different exchange
fields on the $A$ and $B$ sites. There emerges a rich variety of
topologically protected states each of which has a characteristic
spin-valley structure. The single Dirac-cone semimetal is such a state that
one gap is closed while the other three gaps are open, evading the
Nielsen-Ninomiya fermion-doubling problem. We have newly discovered a hybrid
topological insulator named the quantum-spin-quantum-anomalous Hall
insulator, where the quantum anomalous Hall effect occurs at one valley and
the quantum spin Hall effect occurs at the other valley. Along its phase
boundary, single-valley semimetals emerge, where only one of the two valleys
is gapless with degenerated spins. These semimetals are also topologically
protected because they appear in the interface of different topological
insulators. Such a spin-valley dependent physics will be observed by optical
absorption or edge modes.
\end{abstract}

\maketitle


\section{\textbf{Introduction}}

The intrinsic degrees of freedom of an electron are its charge and spin,
which lead to electronics and spintronics. The valley degree of freedom on
honeycomb lattices is expected to provide us with the notion of
valleytronics. Valleytronics was originally proposed in graphene\cite%
{Rycerz,Xiao07,Yao08}, where the states near the Fermi energy are $\pi $
orbitals residing near the $K$ and $K^{\prime }$ points at opposite corners
of the hexagonal Brillouin zone. The low-energy dynamics in the $K$ and $%
K^{\prime }$ valleys is described by the Dirac theory. The valley
excitations are protected by the suppression of intervalley scattering.
However it is hard to realize valleytronics in graphene since the gap is
closed and since it is difficult to discriminate between the $K$ and $%
K^{\prime }$ points experimentally. In this context, transition metal
dichalcogenides\cite{Xiao,Zeng,Cao,Li} become a new playground of
valleytronics, where a considerably large gap is open. In the most recent
experimental progress, the identity of valleys manifests as valley-selective
circular dichroism, leading to valley polarization with circularly polarized
light, offering a possibility to a realization of valleytronics\cite%
{Yao08,Xiao,Zeng,Cao,Li}.

Recently, another honeycomb system of silicon named silicene has been
experimentally synthesized\cite{GLayPRL,Takamura,Takagi} and theoretically
explored\cite{LiuPRL,EzawaNJP,EzawaQAH,EzawaPhoto}. As prominent properties,
it consists of buckled sublattices made of $A$ sites and $B$ sites, and the
Dirac electron has a mass. The buckled structure allows us to control the
Dirac mass independently at the $K$ and $K^{\prime }$ points by applying
external fields such that electric field\cite{EzawaNJP}, exchange field\cite%
{EzawaQAH} and photo-irradiation\cite{EzawaPhoto}. It is possible to
generate a rich variety of topologically protected states in silicene, each
of which has a characteristic spin-valley structure.

In this paper we propose to make a full control of the Dirac mass in order
to search for new topological states together with new spin-valley
structures in silicene. The Dirac mass can be fully controlled by four
potential terms corresponding to the spin and valley degrees of freedom,
among which three terms have already been studied\cite%
{EzawaNJP,EzawaQAH,EzawaPhoto}. The last one is driven by applying the
staggered exchange field $\Delta M\equiv M_{A}-M_{B}$, where $M_{A}$ and $%
M_{B}$ are exchange fields operating on the $A$ and $B$ sites, respectively.
These four terms move Dirac cones, respecting the electron-hole symmetry. It
is also possible to introduce four other potential terms, which shift Dirac
cones so as to break the electron-hole symmetry. The typical one is driven
by applying the mean staggered exchange field $\overline{M}\equiv \frac{1}{2}%
(M_{A}-M_{B})$.

Well-known topologically protected states are quantum spin Hall (QSH)
insulator \cite{KaneMele} and quantum anomalous Hall (QAH) insulator\cite%
{Onoda,Qiao,Tse,Ding,QiaoG,Liu}. They are characterized by the helical and
chiral gapless edge modes, respectively, according to the bulk-edge
correspondence\cite{Hasan,Qi,Wu}. The QAH effect is the quantum Hall effect
without Landau levels, while the QSH effect is the quantum Hall effect of
spins rather than charges.

By introducing the staggered exchange field $\Delta M$, we obtain rich phase
diagrams as illustrated in Fig.1 and Fig.3. First of all, we are able to
generate the spin-polarized QAH (SQAH) insulator together with single
Dirac-cone (SDC) semimetals along its phase boundaries [Fig.2]. The SDC
semimetal is a remarkable state that has one massless Dirac cone and three
massive Dirac cones, evading the Nielsen-Ninomiya fermion-doubling problem%
\cite{Nielsen}. Second, a new finding is the quantum-spin-quantum-anomolous
Hall (QSQAH) insulator. It is a new type of topological insulator such that,
e.g., the QAH effect is realized at the $K$ point while the QSH effect is
realized at the $K^{\prime }$ point [Fig.4]. Third, another new finding is a
single-valley (SV) semimetal such that, e.g., the gap is open (closed) at
the $K$ ($K^{\prime }$) point with spin degeneracy. It is different from the
SDC state which has only one closed gap without spin degeneracy. These
spin-valley dependent band structures will be experimentally observed by
spin-valley selective circular dichroism\cite%
{Yao08,Xiao,Zeng,Cao,Li,EzawaOpt}. We point out that a (semi)metallic state
appearing at the phase boundary between two topologically distinctive
insulators is also protected topologically. We may call it a topological
(semi)mental.

In what follows we use notations $s_{z}=\uparrow \downarrow $, $t_{z}=A,B$, $%
\eta =K,K^{\prime }$ in indices while $s_{z}=\pm 1$, $t_{z}=\pm 1$, $\eta
=\pm 1$ in equations for the spin, the sublattice pseudospin and the valley,
respectively. We also use the Pauli matrices $\sigma _{a}$ and $\tau _{a}$
for the spin and the sublattice pseudospin, respectively.

\section{\textbf{Hamiltonian}}

\label{SecHamil}

Silicene is well described by the tight-binding model\cite{KaneMele,LiuPRB}, 
\begin{eqnarray}
H &=&-t\sum_{\left\langle i,j\right\rangle \alpha }c_{i\alpha }^{\dagger
}c_{j\alpha }+i\frac{\lambda _{\text{SO}}}{3\sqrt{3}}\sum_{\left\langle
\!\left\langle i,j\right\rangle \!\right\rangle \alpha \beta }\nu
_{ij}c_{i\alpha }^{\dagger }\sigma _{\alpha \beta }^{z}c_{j\beta }  \notag \\
&&-i\frac{2}{3}\lambda _{\text{R2}}\sum_{\left\langle \!\left\langle
i,j\right\rangle \!\right\rangle \alpha \beta }t_{z}^{i}c_{i\alpha
}^{\dagger }\left( \mathbf{\sigma }\times \hat{\mathbf{d}}_{ij}\right)
_{\alpha \beta }^{z}c_{j\beta },  \label{KaneMale}
\end{eqnarray}%
where $c_{i\alpha }^{\dagger }$ creates an electron with spin polarization $%
\alpha $ at site $i$ in a honeycomb lattice, and $\left\langle
i,j\right\rangle /\left\langle \!\left\langle i,j\right\rangle
\!\right\rangle $ run over all the nearest/next-nearest-neighbor hopping
sites. The first term represents the usual nearest-neighbor hopping with the
transfer energy $t=1.6$eV. The second term represents the effective
spin-orbit (SO) interaction with $\lambda _{\text{SO}}=3.9$meV, and $\nu
_{ij}=+1$ if the next-nearest-neighboring hopping is anticlockwise and $\nu
_{ij}=-1$ if it is clockwise with respect to the positive $z$ axis. The
third term represents the Rashba interaction with $\lambda _{\text{R2}}=0.7$%
meV, where $t_{z}^{i}=\pm 1$ for $i$ representing the $A$ ($B$) site; $\hat{%
\mathbf{d}}_{ij}=\mathbf{d}_{ij}/\left\vert \mathbf{d}_{ij}\right\vert $
with the vector $\mathbf{d}_{ij}$ connecting two sites $i$ and $j$ in the
same sublattice.

The low-energy effective Hamiltonian is given by the Dirac theory around the 
$K_{\eta }$ point. The Hamiltonian (\ref{KaneMale}) yields%
\begin{eqnarray}
H_{\eta }^{0} &=&\hbar v_{\text{F}}\left( \eta k_{x}\tau _{x}+k_{y}\tau
_{y}\right) +\lambda _{\text{SO}}\eta \tau _{z}\sigma _{z}  \notag \\
&&+a\lambda _{\text{R2}}\eta \tau _{z}\left( k_{y}\sigma _{x}-k_{x}\sigma
_{y}\right) ,  \label{DiracHamil}
\end{eqnarray}%
where $v_{\text{F}}=\frac{\sqrt{3}}{2}at$ is the Fermi velocity with the
lattice constant $a=3.86$\AA .

A great merit of silicene is that we can introduce various potential terms
into the Hamiltonian by making advantages of its bucked structure. There
exist eight commuting terms which we are able to introduce into the Dirac
Hamiltonian (\ref{DiracHamil}). They are%
\begin{equation}
H_{pqr}=\lambda _{pqr}\eta ^{p}(\sigma _{z})^{q}(\tau _{z})^{r},
\label{pqrTerm}
\end{equation}%
where $p,q,r=0$ or $1$.

The coefficient of $\tau _{z}$ is the Dirac mass, to which four terms
contribute. They are $H_{pqr}$ with $r=1$. First, $H_{111}$ is nothing but
the SO coupling term with $\lambda _{111}=\lambda _{\text{SO}}$. Second, $%
H_{001}$ is the staggered sublattice potential term\cite{EzawaNJP} with $%
\lambda _{001}=\ell E_{z}$, which is controlled by applying electric field $%
E_{z}$, where $\ell =0.23$\r{A}\ is the sublattice separation. Third, $%
H_{101}$ is the Haldane term\cite{Haldane} with strength $\lambda
_{101}=\lambda _{\Omega }$, which is controlled by applying photo-irradiation%
\cite{EzawaPhoto}. Finally, $H_{011}$ is the new term we propose to analyze
in the present work. As we see, it is the staggered exchange term with $%
\lambda _{011}=\Delta M$. We summarize the property of the Dirac mass term $%
H_{pq1}$ with respect to the time-reversal symmetry (TRS), the spin-rotation
symmetry (SRS) and the sublattice pseudospin symmetry (SLS)\cite{Ryu} as 
\begin{equation}
\begin{tabular}{||c|c|c|c|c||}
\hline\hline
$H_{pq1}$ & potential term & TRS & SRS & SLS \\ \hline
111 & Kane-Mele & true & false & false \\ \hline
001 & s-sublattice & true & true & false \\ \hline
011 & s-exchange & false & false & false \\ \hline
101 & Haldane & false & true & false \\ \hline\hline
\end{tabular}%
\,,
\end{equation}%
where s- stands for staggered. We find that QAH effects can be induced by
the Haldane term or the staggered exchange term since they break the TRS.

We address the other potential terms in (\ref{pqrTerm}), or $H_{pqr}$ with $%
r=0$. They induce the shift of Dirac cones and break the electron-hole
symmetry, as we discuss in Section \ref{SecMdM}. First, $H_{000}$ is nothing
but the chemical potential with $\lambda _{000}=\mu $. Second, $H_{010}$ is
the mean exchange coupling term\cite{EzawaQAH} with $\lambda _{010}=%
\overline{M}$. The rest two terms $H_{100}$ and $H_{110}$ have not been
discussed previously, and their experimental realizations would yet be
explored. We may call $H_{100}$ the staggered Haldane term with $\lambda
_{100}=\lambda _{\text{SH}}$ and $H_{110}$ the staggered Kane-Mele term with 
$\lambda _{110}=\lambda _{\text{SKM}}$ . We summarize the symmetry property
of the shift term $H_{pq1}$ as

\begin{equation}
\begin{tabular}{||c|c|c|c|c||}
\hline\hline
$H_{pq0}$ & potential term & TRS & SRS & SLS \\ \hline
000 & chemical potential & true & true & true \\ \hline
010 & m-exchange & false & false & true \\ \hline
100 & s-Haldane & false & true & true \\ \hline
110 & s-Kane-Mele & true & false & true \\ \hline\hline
\end{tabular}%
\,,
\end{equation}%
where m- and s- stands for mean and staggered, respectively.

We may write down the tight-binding term that yields the potential term $%
H_{pqr}$. The additional terms are\cite{EzawaNJP,EzawaQAH}%
\begin{align}
\Delta H=& -\ell \sum_{i\alpha }t_{z}^{i}E_{z}c_{i\alpha }^{\dagger
}c_{i\alpha }+i\frac{\lambda _{\Omega }}{3\sqrt{3}}\sum_{\left\langle
\!\left\langle i,j\right\rangle \!\right\rangle \alpha \beta }\nu
_{ij}c_{i\alpha }^{\dagger }c_{j\beta }  \notag \\
& +\sum_{i\alpha }M_{t_{z}^{i}}c_{i\alpha }^{\dagger }\sigma _{\alpha \alpha
}^{z}c_{i\alpha }+\mu \sum_{i\alpha }c_{i\alpha }^{\dagger }c_{i\alpha } 
\notag \\
& +i\frac{\lambda _{\text{SH}}}{3\sqrt{3}}\sum_{\left\langle \!\left\langle
i,j\right\rangle \!\right\rangle \alpha \beta }\tau _{z}\nu _{ij}c_{i\alpha
}^{\dagger }c_{j\alpha }  \notag \\
& +i\frac{\lambda _{\text{SKM}}}{3\sqrt{3}}\sum_{\left\langle \!\left\langle
i,j\right\rangle \!\right\rangle \alpha \beta }\tau _{z}\nu _{ij}c_{i\alpha
}^{\dagger }\sigma _{\alpha \beta }^{z}c_{j\beta }.  \label{BasicHamil}
\end{align}%
The Dirac Hamiltonian is%
\begin{eqnarray}
H_{\eta } &=&H_{\eta }^{0}-\ell E_{z}\tau _{z}+\eta \lambda _{\Omega }\tau
_{z}+\Delta M\sigma _{z}\tau _{z}  \notag \\
&&+\mu +\overline{M}\sigma _{z}+\lambda _{\text{SH}}\eta +\lambda _{\text{SKM%
}}\eta \sigma _{z},  \label{TotalDirac}
\end{eqnarray}%
where $\overline{M}=\frac{1}{2}(M_{A}+M_{B})$ and $\Delta M=M_{A}+M_{B}$.
The Dirac mass is given by%
\begin{equation}
\Delta _{s_{z}}^{\eta }=\eta s_{z}\lambda _{\text{SO}}-\ell E_{z}+\eta
\lambda _{\Omega }+s_{z}\Delta M,  \label{DiracMass}
\end{equation}%
which may be positive, negative or zero. We can make a full control of the
Dirac mass independently for each spin and valley to materialize
spin-valleytronics in silicene.

\begin{figure}[t]
\centerline{\includegraphics[width=0.30\textwidth]{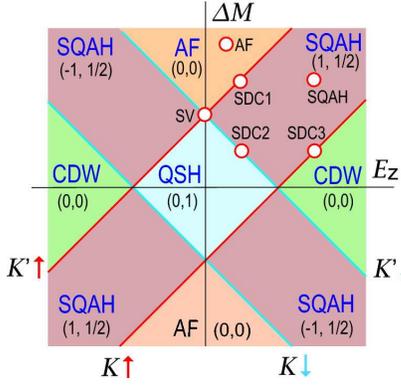}}
\caption{(Color online) Phase diagram in the $\ell E_{z}$-$\Delta M$ plane.
Heavy lines represent phase boundaries where the band gap closes. There are
three types of topological insulators as indicated by QSH with $(0,1)$ and
SQAH with $(\pm 1,\frac{1}{2})$. There are two types of trivial band
insulators as indicated by AF and CDW. There emerge the SDC semimetal and
the SV semimetal in the phase boundary. A circle shows a point where the
energy spectrum is shown in Fig.\protect\ref{FigMEBand}. The gap closes at $%
E_{z}=\pm 17$meV/\AA\ on the $E_{z}$ axis and at $\Delta M=\pm \protect%
\lambda _{\text{SO}}$ on the $\Delta M$ axis.}
\label{FigPhaseME}
\end{figure}

\begin{figure}[t]
\centerline{\includegraphics[width=0.5\textwidth]{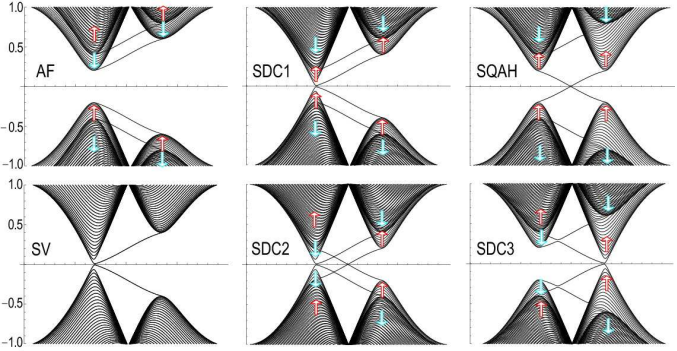}}
\caption{(Color online) The band structures of zigzag silicene nanoribbons
at the points indicated in the phase diagram [Fig.\protect\ref{FigPhaseME}].
The vertical axis is the energy in unit of $t$, and the horizontal axis is
the momentum. We can clearly see the Dirac cones representing the energy
spectrum of the bulk. Lines connecting the two Dirac cones are edge modes.}
\label{FigMEBand}
\end{figure}

\section{\textbf{Topological charges}}

\label{SecTC}

We first explore the Hamiltonian system with the mass correction terms $%
H_{pq1}$ included. The Hamiltonian $H_{\eta }$\ is explicitly given by the $%
4\times 4$ matrix as%
\begin{equation}
H_{\eta }=\left( 
\begin{array}{cc}
H_{\eta }^{\uparrow } & R_{\eta } \\ 
R_{\eta }^{\dag } & H_{\eta }^{\downarrow }%
\end{array}%
\right) ,  \label{BasicHamil4}
\end{equation}%
with the diagonal elements%
\begin{equation}
H_{\eta }^{s_{z}}=\left( 
\begin{array}{cc}
\Delta _{s_{z}}^{\eta } & \hbar v_{\text{F}}(\eta k_{x}-ik_{y}) \\ 
\hbar v_{\text{F}}(\eta k_{x}+ik_{y}) & -\Delta _{s_{z}}^{\eta }%
\end{array}%
\right) ,
\end{equation}%
and the off-diagonal element%
\begin{equation}
R_{\eta }=\left( 
\begin{array}{cc}
ia\lambda _{\text{R2}}(\eta k_{x}-ik_{y}) & 0 \\ 
0 & -ia\lambda _{\text{R2}}(\eta k_{x}-ik_{y})%
\end{array}%
\right) .  \label{HamilR}
\end{equation}%
Note that the off-diagonal element $R_{\eta }$\ vanishes at the $K_{\eta }$
point where $k_{x}=k_{y}=0$.

The characteristic feature is the existence of the electron-hole symmetry.
The energy spectrum at the $K_{\eta }$ point contains four levels, $\pm
\Delta _{s_{z}}^{\eta }$, and the band gap is given by $2|\Delta
_{s_{z}}^{\eta }|$. Topological phase transitions are controlled entirely by
the spin-valley dependent Dirac mass $\Delta _{s_{z}}^{\eta }$.

We set $\lambda _{\text{R2}}=0$ since it is a small quantity. We are able to
justify this simplification in the present system, as we describe at the end
of this section. When $\lambda _{\text{R2}}=0$, the spin $s_{z}$ is a good
quantum number. The Dirac Hamiltonian is given by the $2\times 2$ matrix $%
H_{\eta }^{s_{z}}$ for each spin $s_{z}$ and each Dirac valley $K_{\eta }$.
For such a system it is straightforward to calculate the spin-valley
dependent Chern number $\mathcal{C}_{s_{z}}^{\eta }$ by integrating the
Berry curvature over all occupied states of electrons in the momentum space.
The Berry curvature is described by the meron configuration in the
sublattice-pseudospin space, and the Chern number $\mathcal{C}_{s_{z}}^{\eta
}$ becomes identical to the Pontryagin number\cite{EzawaEPJB}. We find%
\begin{equation}
\mathcal{C}_{s_{z}}^{\eta }={\frac{\eta }{2}}\text{sgn}(\Delta
_{s_{z}}^{\eta })  \label{ChernNumbe}
\end{equation}%
as a function of the spin-valley dependent Dirac mass $\Delta _{s_{z}}^{\eta
}$. This is well defined provided that the Fermi level is taken within the
insulating gap. A phase transition may occur when one Dirac mass becomes
zero, $\Delta _{s_{z}}^{\eta }=0$ for certain $\eta $ and $s_{z}$.

The topological quantum numbers are the Chern number $\mathcal{C}$ and the
spin-Chern number $\mathcal{C}_{s}$ modulo $2$. They are given by\beginABC%
\label{TopNumDM}

\begin{align}
\mathcal{C}& =\mathcal{C}_{\uparrow }^{K}+\mathcal{C}_{\uparrow }^{K^{\prime
}}+\mathcal{C}_{\downarrow }^{K}+\mathcal{C}_{\downarrow }^{K^{\prime }}, \\
\mathcal{C}_{s}& =\frac{1}{2}\left( \mathcal{C}_{\uparrow }^{K}+\mathcal{C}%
_{\uparrow }^{K^{\prime }}-\mathcal{C}_{\downarrow }^{K}-\mathcal{C}%
_{\downarrow }^{K^{\prime }}\right) .
\end{align}%
\endABC We now switch on the Rashba interaction adiabatically, $\lambda _{%
\text{R2}}\neq 0$. As far as $\lambda _{\text{R2}}$ is small, the band
structure is almost unchanged. Furthermore, it follows from the Hamiltonian (%
\ref{BasicHamil4}) that the phase transition point is independent of $%
\lambda _{\text{R2}}$ and still given by solving $\Delta _{s_{z}}^{\eta }=0$%
. Since the gap keeps open during this adiabatic process, the Chern numbers
are well defined and their values are unchanged since they are quantized.

\section{\textbf{Phase diagram in }$(E_{z},\Delta M)$\textbf{\ plane}}

\label{SecPhaseEM}

For definiteness we investigate the topological phase transition in the $%
(E_{z},\Delta M)$ plane, where%
\begin{equation}
\Delta _{s_{z}}^{\eta }=\eta s_{z}\lambda _{\text{SO}}-\ell
E_{z}+s_{z}\Delta M.
\end{equation}%
The phase boundaries are given by solving $\Delta _{s_{z}}^{\eta }=0$, which
yields four heavy lines corresponding to $s_{z}=\uparrow \downarrow $ and $%
\eta =K,K^{\prime }$ in the phase diagram [Fig.\ref{FigPhaseME}]. The
spin-valley dependent Chern number $\mathcal{C}_{s_{z}}^{\eta }$ is
calculated at each point $(E_{z},\Delta M)$ with the use of (\ref{ChernNumbe}%
), from which we derive the topological numbers $(\mathcal{C}$, $\mathcal{C}%
_{s})$ based on (\ref{TopNumDM}). They take constant values in one phase,
which we have depicted in the phase diagram. We illustrate the band
structure of a nanoribbon with zigzag edges in Fig.\ref{FigMEBand}, which
manifests the spin-valley structure of topologically protected phases.

First, there appear four types of insulators:

(1) The spin-polarized QAH (SQAH) insulators with $(\mathcal{C},\mathcal{C}%
_{s})=(\pm 1,\frac{1}{2})$.

(2) The QSH insulator with $(0,1)$.

(3) The trivial charge-density-wave-type (CDW) band insulator with $(0,0)$.

(4) The trivial antiferromagnetic-order-type (AF) band insulator with $(0,0)$.

We note that there are two types of trivial band insulators. The band gaps
are different between the $K$ and $K^{\prime }$ points in the AF insulator,
while they are identical in the CDW insulator.

Second, SDC metals appear in the three phase boundaries of the SQAH
insulator. The SDC metal is originally found in silicene by applying
photo-irradiation and electric field simultaneously\cite{EzawaPhoto}. It is
interesting that the SDC state is also obtainable without photo-irradiation.
On the other hand, the SV semimetal appears at the point where the four
topological insulators meets. They are topologically protected semimetals,
since they appear in the interface of different topological insulators each
of which is topologically protected against small perturbations.

\section{Inhomogeneous Dirac mass}

We may apply an inhomogeneous electric field\cite{EzawaNJP} $E_{z}\left(
x,y\right) $ or generate a domain wall in the antiferromagnet $\Delta
M\left( x,y\right) $, which makes the Dirac mass inhomogeneous. For
simplicity we assume the homogeneity in the $x$ direction. The zero modes
appear along the line determined by $\Delta _{s_{z}}^{\eta }\left( y\right)
=0$, when $\Delta _{s_{z}}^{\eta }\left( y\right) $ changes the sign. We may
set $k_{x}=$constant due to the translational invariance along the $x$ axis.
We seek the zero-energy solution. The particle-hole symmetry guarantees the
existence of zero-energy solutions satisfying the relation $\psi _{B}=i\xi
\psi _{A}$ with $\xi =\pm 1$. Here, $\psi _{A}$ is a two-component amplitude
with the up spin and down spin. Setting $\psi _{A}\left( x,y\right)
=e^{ik_{x}x}\phi _{A}\left( y\right) $, we obtain $H_{\eta }\psi _{A}\left(
x,y\right) =E_{\eta \xi }\psi _{A}\left( x,y\right) $, together with a
linear dispersion relation $E_{\eta \xi }=\eta \xi \hbar v_{\text{F}}k_{x}$.
We can explicitly solve this as 
\begin{equation}
\phi _{As_{z}}\left( y\right) =C\exp \left[ \frac{\xi }{\hbar v_{\text{F}}}%
\int^{y}\Delta _{s_{z}}^{\eta }\left( y^{\prime }\right) dy^{\prime }\right]
,  \label{ZeroModeSolut}
\end{equation}%
where $C$ is the normalization constant. The sign $\xi $ is determined so as
to make the wave function finite in the limit $\left\vert y\right\vert
\rightarrow \infty $. This is a reminiscence of the Jackiw-Rebbi mode\cite%
{Jakiw} presented for the chiral mode. The difference is the presence of the
spin and valley indices in the wave function.

\begin{figure}[t]
\centerline{\includegraphics[width=0.30\textwidth]{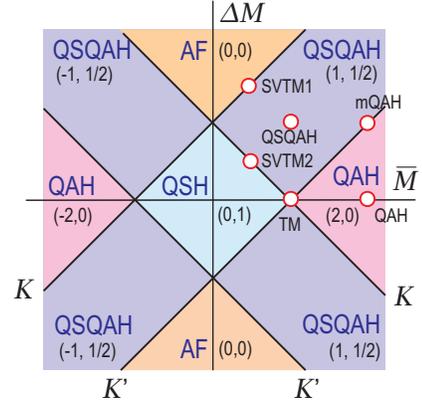}}
\caption{(Color online) Phase diagram in the $\overline{M}$-$\Delta M$
plane. Heavy lines represent phase boundaries. There are five types of
topological insulators as indicated by QAH with $(\pm 2,0)$, QSQAH with $%
(\pm 1,1/2)$ and QSH with $(0,1)$. There is one trivial band insulator as
indicated by AF. There emerge the mQAH metal and the TM in the phase
boundary. A circle shows a point where the energy spectrum is calculated and
shown in Fig.\protect\ref{FigBandMdM}. The gap closes at $\overline{M}=\pm 
\protect\lambda_{\text{SO}}$ on the $\overline{M}$ axis and at $\Delta M=\pm 
\protect\lambda_{\text{SO}}$ on the $\Delta M$ axis.}
\label{FigPhaseMdM}
\end{figure}

\begin{figure}[t]
\centerline{\includegraphics[width=0.50\textwidth]{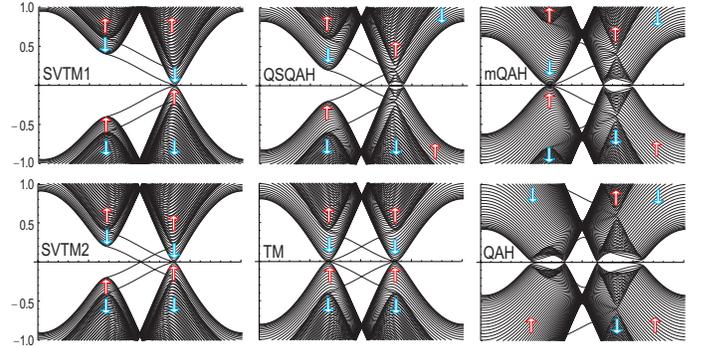}}
\caption{(Color online) The band structures of silicene nanoribbons at
marked points in the phase diagram [Fig.\protect\ref{FigPhaseMdM}]. The
vertical axis is the energy in unit of $t$, and the horizontal axis is the
momentum. Lines connecting the two Dirac cones are edge modes.}
\label{FigBandMdM}
\end{figure}

\section{\textbf{Phase diagram in }$(\overline{M},\Delta M)$\textbf{\ plane}}

\label{SecMdM}

We next investigate the Hamiltonian system by including the Dirac-cone
shifting terms $H_{pq0}$. General analysis is quite difficult to make since
there exists no electron-hole symmetry. For definiteness we only consider
the mean-exchange-field term $H_{010}$. The Hamiltonian is given by (\ref%
{BasicHamil4}) together with%
\begin{equation}
H_{\eta }^{s_{z}}=\left( 
\begin{array}{cc}
\Delta _{s_{z}}^{\eta }+\overline{M}s_{z} & \hbar v_{\text{F}}(\eta
k_{x}-ik_{y}) \\ 
\hbar v_{\text{F}}(\eta k_{x}+ik_{y}) & -\Delta _{s_{z}}^{\eta }+\overline{M}%
s_{z}%
\end{array}%
\right) ,  \label{HamilX}
\end{equation}%
where the Dirac mass $\Delta _{s_{z}}^{\eta }$ is given by (\ref{DiracMass}%
). The gap is no longer given solely by the Dirac mass.

The energy spectrum at the $K_{\eta }$ point contains four levels, $\Delta
_{s_{z}}^{\eta }+\overline{M}s_{z}$ and $-\Delta _{s_{z}}^{\eta }+\overline{M%
}s_{z}$ with $s_{z}=\pm 1$. The gap closes at the $K_{\eta }$ point when any
two of them coincide,%
\begin{equation}
\Delta _{s_{z}}^{\eta }=\pm \overline{M}s_{z},  \label{PhaseBound}
\end{equation}%
which in general gives the phase boundaries. However, the present analysis
is not enough to construct the phase diagram, because the Dirac cones with
the opposite spins touch one to another at the phase transition point. When
they touch, the Rashba interaction ($\lambda _{\text{R2}}\neq 0$) mixes the
up and down spins, yielding an essential modification of the band structure%
\cite{EzawaQAH}. It is necessary to diagonalize the Hamiltonian (\ref%
{BasicHamil4}) with the off-diagonal element (\ref{HamilR}) and the diagonal
element (\ref{HamilX}).

Let us study explicitly the simplest case with the exchange fields $M_{A}$
and $M_{B}$ without any other external fields, where%
\begin{equation}
\Delta _{s_{z}}^{\eta }=\eta s_{z}\lambda _{\text{SO}}+s_{z}\Delta M.
\end{equation}%
We give the phase diagram in the $(\overline{M},\Delta M)$ plane [Fig.\ref%
{FigPhaseMdM}]. The phase boundaries are given by solving (\ref{PhaseBound}%
), which yield four heavy lines corresponding to $s_{z}=\uparrow \downarrow $
and $\eta =K,K^{\prime }$ in the phase diagram. We have four distinct
regions referred to as QSH, AF, QAH and QSQAH. We also illustrate the band
structure of a silicene nanoribbon with zigzag edges [Fig.\ref{FigBandMdM}]
for typical points in the phase diagram.

When we set $\lambda _{\text{R2}}=0$, only the QSH phase is an insulator. In
all the other regions the conduction and valence bands penetrate into one
another and are mixed.

When we include the Rashba interaction ($\lambda _{\text{R2}}\neq 0$), the
collision is avoided since it turns the level crossing into the level
anticrossing. Indeed, the energy spectrum consists of four levels,%
\begin{equation}
E_{s_{z}}^{\eta }=\pm \sqrt{(a\lambda _{\text{R2}}k)^{2}+F_{\pm }(k)^{2}},
\end{equation}%
with $F_{\pm }(k)=\eta s_{z}\overline{M}\pm \sqrt{(\hbar v_{\text{F}%
}k)^{2}+\left( \eta \lambda _{\text{SO}}+\Delta M\right) ^{2}}$. We can see
that the gap is open everywhere in the phase diagram except for the phase
boundaries: Some typical examples of nanoribbon band structure are found in
Fig.\ref{FigBandMdM}.

Since the spin is no longer a good quantum number, the topological numbers
are no longer given by the formulas (\ref{TopNumDM}). The Berry curvature is
described by Skyrmions in the real spin space\cite{EzawaQAH}. We are able to
introduce the pseudospin-valley Chern number $\mathcal{C}_{t_{z}}^{\eta }$
with $t_{z}=A,B$,%
\begin{equation}
\mathcal{C}_{t_{z}}^{\eta }=\frac{1}{2}\text{sgn}\left( \eta t_{z}\lambda _{%
\text{SO}}+M_{t_{z}}\right) ,
\end{equation}%
in terms of which the topological numbers are given by\beginABC%
\begin{eqnarray}
\mathcal{C} &=&\mathcal{C}_{A}^{K}+\mathcal{C}_{B}^{K}+\mathcal{C}%
_{A}^{K^{\prime }}+\mathcal{C}_{B}^{K^{\prime }}, \\
\mathcal{C}_{s} &=&\frac{1}{2}\left( C_{A}^{K}-C_{B}^{K}-C_{A}^{K^{\prime
}}+C_{B}^{K^{\prime }}\right) .
\end{eqnarray}%
\endABC They are calculated at each point $(M_{A},M_{B})$ or $(\overline{M}%
,\Delta M)$. The topological numbers $(\mathcal{C},\mathcal{C}_{s})$ take
constant values in one phase, which we have depicted in the phase diagram
[Fig.\ref{FigPhaseMdM}]. We illustrate the band structure of a nanoribbon
with zigzag edges in Fig.\ref{FigBandMdM}. Note that the spin becomes almost
a good quantum number away from the $K_{\eta }$ points.

First, there appear four types of insulators:

(1) The QSQAH insulators with $(C,C_{s})=(\pm 1,1/2)$, where the QAH effect
is realized at one valley in coexistence with the QSH effect at the other
valley. It is intriguing that, e.g., the topological numbers $(\pm 1,0)$ are
assigned to the $K$ point and $(0,1/2)$ to the $K^{\prime }$ point.

(2) The trivial AF insulator with $(0,0)$.

(3) The QSH insulator with $(0,1)$.

(4) The QAH phases with $(2,0)$ for $M>0$ and $(-2,0)$ for $M<0$.

Second, at the boundary, the topological metal (TM) and the single-valley
topological metal (SVTM) emerge. The band touches parabolically in them.
Their emergencies are also protected topologically since they are sandwiched
by different topological insulators.

\section{\textbf{Optical absorption}}

An interesting experiment to probe and manipulate the valley degree of
freedom is optical absorption\cite{Yao08,Xiao,Zeng,Cao,Li,EzawaOpt}. We
briefly discuss that spin-valley dependent band structures we have found
will be experimentally observable by spin-valley-selective circular
dichroism. Circular dichroism is a phenomena in which the response of the
left- and right-handed circularly polarized light is different. To assess
the optical selectivity of spin and valley by circularly polarized light, we
compute the spin-valley dependent degree of circular polarization between
the top valence bands and bottom of conduction bands. It is straightforward
to apply the standard method\cite{Yao08,Xiao,Zeng,Cao,Li,EzawaOpt} to
calculate an optical absorption.

We consider the interband matrix elements of the left- and right-polarized
radiation fields ($\pm $) for spin $s_{z}$ at $k$ for a vertical transition
from band $u_{\text{c}}$ to band $u_{\text{v}}$. They are defined by%
\begin{equation}
P_{\pm }^{\eta }\left( k\right) \equiv m_{0}\left\langle u_{\text{c}}\left(
k\right) \right\vert \frac{1}{\hbar }\frac{\partial H_{\eta }}{\partial
k_{\pm }}\left\vert u_{\text{v}}\left( k\right) \right\rangle .
\end{equation}%
If we neglect the Rashba terms ($\lambda _{\text{R2}}=0$), we are able to
obtain an analytic formula for the transitions near the $K_{\eta }$ point as%
\begin{equation}
\left\vert P_{\pm }^{\eta }\left( k\right) \right\vert ^{2}=m_{0}^{2}v_{%
\text{F}}^{2}\left( 1\pm \eta \frac{\Delta _{s_{z}}^{\eta }}{\sqrt{\left(
\Delta _{s_{z}}^{\eta }\right) ^{2}+4a^{2}t^{2}k^{2}}}\right) ^{2}.
\end{equation}%
Especially we find the spin-valley dependent optical selection rule at $k=0$,%
\begin{equation}
\left\vert P_{\pm }^{\eta }\left( 0\right) \right\vert ^{2}=m_{0}^{2}v_{%
\text{F}}^{2}[1\pm \eta \text{sgn}\left( \Delta _{s_{z}}^{\eta }\right)
]^{2},
\end{equation}%
which will be detected experimentally. Such a spin-valley selective circular
dichroism would lead to the eventual realization of spin-valleytronics.

\section{Discussion}

In this paper, exploiting advantages of the buckled structure, we have
proposed to make a full control of the Dirac mass and hence a full control
of the topological charges in silicene. A topological state has a particular
spin-valley structure. In exploring the phase diagram, we have found a
hybrid topological insulator named Quantum-Spin-Quantum-Anomalous Hall
(QSQAH) insulator, where two different topological insulators coexist: The
QSH effect is realized at one valley while the QAH effect is realized at the
other valley. The topological numbers are simply given by one half of the
sum of those of the QSH and QAH insulators. Such a hybrid of two distinctive
topological insulators is utterly a new state.

We address a question how realistic it is to generate various topological
insulators such as the QSQAH state. In order to make the QSQAH phase, the
exchange interaction at the A (B) site should be larger (smaller) than the
SO interaction, whose magnitude is $3.9$meV. When we attach a ferromagnet to
the side of A sites, the exchange field at B sites is naturally smaller than
that at A sites. Since the magnitude of the exchange field induced by
transition metal deposition is of the same order of the SO interaction\cite%
{Qiao}, it is plausible that the condition for the QSQAH phase can be
satisfied in a realistic experimental situation. Furthermore, we can think
of three possible methods to produce exchange fields acting separately on
the A and B sublattices. First, we sandwich silicene by two different
ferromagnets. Second, we arrange two different transition metals so as to be
absorbed as adatoms separately to the A and B sublattices. Third, we attach
a honeycomb-lattice antiferromagnet\cite{Li} or ferrimagnet to silicene.
What kind of materials one should use will be a future problem.

I am very much grateful to N. Nagaosa for many fruitful discussions on the
subject. This work was supported in part by Grants-in-Aid for Scientific
Research from the Ministry of Education, Science, Sports and Culture No.
22740196.


\begin{thebibliography}{99}
\bibitem{Rycerz} A. Rycerz, J. Tworzydlo, and C. W. J. Beenakker, Nat,\quad
Phys. \textbf{3}, 172 (2007).

\bibitem{Xiao07} D. Xiao, W. Yao, and Q. Niu, Phys. Rev. Lett. \textbf{99},
236809 (2007).

\bibitem{Yao08} W. Yao, D. Xiao, and Q. Niu, Phys. Rev. B \textbf{77},
235406 (2008).

\bibitem{Xiao} D. Xiao, G.-B. Liu, W. Feng, X. Xu, and W. Yao, Phys. Rev.
Lett. \textbf{108}, 196802 (2012).

\bibitem{Zeng} H. Zeng, J. Dai, W. Yao, D. Xiao and X. Cui, Nat. Nanotech. 
\textbf{7}, 490 (2012).

\bibitem{Cao} T. Cao, G. Wang, W. Han, H. Ye, C. Zhu, J. Shi, Q. Niu, P.
Tan, E. Wang, B. Liu and J Feng Nat. Com. \textbf{3}, 887 (2012).

\bibitem{Li} X. Li, T. Cao, Q. Niu, J. Shi, and J Feng,
cond-mat/arXiv:1210.4623

\bibitem{GLayPRL} P. Vogt, P. De Padova, C. Quaresima, J. A., E.
Frantzeskakis, M. C. Asensio, A. Resta, B. Ealet and G. L. Lay, Phys. Rev.
Lett. \textbf{108}, 155501 (2012).

\bibitem{Takamura} A. Fleurence, R. Friedlein, T. Ozaki, H. Kawai, Y. Wang,
and Y. Yamada-Takamura, Phys. Rev. Lett. \textbf{108}, 245501 (2012).

\bibitem{Takagi} C.-L. Lin, R. Arafune, K. Kawahara, N. Tsukahara, E.
Minamitani, Y. Kim, N. Takagi, M. Kawai, Appl. Phys. Express \textbf{5},
045802 (2012).

\bibitem{LiuPRL} C.-C. Liu, W. Feng, and Y. Yao, Phys. Rev. Lett. \textbf{107%
}, 076802 (2011).

\bibitem{EzawaNJP} M. Ezawa, New J. Phys. \textbf{14}, 033003 (2012).

\bibitem{EzawaQAH} M. Ezawa, Phys. Rev. Lett. \textbf{109}, 055502 (2012).

\bibitem{EzawaPhoto} M. Ezawa, Phys. Rev. Lett. \textbf{110}, 026603 (2013).

\bibitem{KaneMele} C. L. Kane and E. J. Mele, Phys. Rev. Lett. \textbf{95},
226801 (2005).

\bibitem{Onoda} M. Onoda and N. Nagaosa, Phys. Rev. Lett. \textbf{90},
206601 (2003).

\bibitem{Qiao} Z. Qiao, S. A. Yang, W. Feng, W-K. Tse, J. Ding, Y. Yao, J.
Wang, and Q. Niu, Phys. Rev. B \textbf{82}, 161414 R (2010).

\bibitem{Tse} W.-K. Tse, Z. Qiao, Y. Yao, A. H. MacDonald, and Q. Niu, Phys.
Rev. B \textbf{83}, 155447 (2011).

\bibitem{Ding} J. Ding, Z. Qiao, W. Feng, Y. Yao and Q. Niu, Phys. Rev. B 
\textbf{84}, 195444 (2011).

\bibitem{QiaoG} Z. Qiao, H. Jiang, X. Li, Y. Yao and Q. Niu, Phys. Rev. B 
\textbf{85}, 115439 (2012).

\bibitem{Liu} C.-X. Liu, X.-L. Qi, X. Dai, Z. Fang and S.-C. Zhang, Phys.
Rev. Lett. \textbf{101}, 146802 (2008).

\bibitem{Hasan} M.Z Hasan and C. Kane, Rev. Mod. Phys. \textbf{82}, 3045
(2010).

\bibitem{Qi} X.-L. Qi and S.-C. Zhang, Rev. Mod. Phys. \textbf{83}, 1057
(2011).

\bibitem{Wu} C. Wu, B.A. Bernevig and S.-C. Zhang, Phys. Rev. Lett. \textbf{%
96}, 106401 (2006).

\bibitem{Nielsen} H. B. Nielsen and M. Ninomiya, Nucl. Phys. B \textbf{185},
20 (1981).

\bibitem{EzawaOpt} M. Ezawa, Phys. Rev. B \textbf{86}, 161407(R) (2012).

\bibitem{LiuPRB} C.-C. Liu, H. Jiang, and Y. Yao, Phys. Rev. B, \textbf{84},
195430 (2011).

\bibitem{Haldane} F. D. M. Haldane, Phys. Rev. Lett. \textbf{61}, 2015
(1988).

\bibitem{Ryu} S. Ryu, C. Mudry, C.Y. Hou and C. Chamon, Phys. Rev. B \textbf{%
80}, 205319 (2009).

\bibitem{EzawaEPJB} M. Ezawa, Eur. Phys. J. B \textbf{85}, 363 (2012).

\bibitem{Jakiw} R. Jackiw and C. Rebbi, Phys. Rev. D \textbf{13}, 3398
(1976).
\end{thebibliography}
\end{document}